\documentclass[]{article}
\usepackage[a4paper, margin=1in]{geometry}
\usepackage{authblk}	
\usepackage{amsmath}    
\usepackage{graphicx}	
\usepackage{caption}	
\usepackage{subcaption}	
\usepackage[flushleft]{threeparttable}	
\usepackage{multirow}	

\usepackage[
	backend=bibtex,
	style=phys,
	sorting=none
]{biblatex}
\title{Intermittency analysis of $pp$ collisions at $\sqrt{s}=$ 0.9, 7 and 8 TeV from the CMS experiment}

\author{Z. Ong, P. Agarwal, H.W. Ang, A.H. Chan, C.H. Oh}
\affil{Department of Physics, National University of Singapore}
\date{}

\begin{document}

\maketitle

\begin{abstract}
	The method of horizontal scaled factorial moments as outlined by Bialas and Peschanski was used to conduct intermittency analysis for $pp$ collisions at $\sqrt{s}=$ 0.9, 7 and 8 TeV from the CMS experiment. The data was obtained and processed from the CERN Open Data Portal. It was found from 1D analysis that the intermittency strength decreases with increasing energy, indicating that the signature of the $\alpha$-model of random cascading that the former is based on seems to be weakening. Intermittency was stronger in 2D, but did not reveal any clear trend with increasing collision energy.
\end{abstract}

\section{Introduction}

Motivated by the observations of spikes and voids in the (pseudo)rapidity distribution of some high energy events from JACEE~\cite{Burnett:1983pb} and NA22~\cite{Adamus:1986vd}) in the late 1980s, Bialas and Peschanski~\cite{Bialas:1985jb,Bialas:1988wc} established the context and laid the foundations for intermittency in multiparticle production, which would go on to engage high energy physicists for more than a decade. Based on the $\alpha$-model of turbulence~\cite{Bialas:1985jb}, they proposed studying this phenomena by a measurement of the bin-averaged scaled factorial moments,
\begin{equation}
	\label{eq:HorizontalMoments}
	F_{q}^{\text{h}}(\delta \eta) = M^{q-1} \sum_{m=1}^{M} \frac{\langle n_{m}(n_{m}-1) \ldots (n_{m}-q+1) \rangle}{\langle N \rangle^{q}},
\end{equation}
where $F_q$ is the $q^{\text{th}}$ bin-averaged moment, $M$ is the number of bins that the (pseudo)rapidity
space is divided into (each with size $\delta \eta$), $n_m$ is the multiplicity in bin $m$, $N$ is the total event
multiplicity and $\left< \dots \right>$ represents an average over events.

Equation~\ref{eq:HorizontalMoments} has come to be known as the \textit{horizontal moments}\footnote{Other variants of the scaled factorial moments have been proposed as well, such as the vertical moments~\cite{Bialas:1985jb,Bialas:1988wc}, which have been found to give unstable results as the bin window decreases (e.g. $\delta \eta \rightarrow 0$). Hence, they will not be used in this analysis.}. The system is considered intermittent if we have
\begin{equation}
	F_{q}(\delta \eta) \propto \left( \frac{1}{\delta\eta}\right)^{\phi_{q}} 
	\label{eq:F-moment-scaling-law}
\end{equation}
as $\delta \eta \rightarrow 0$. $\phi_{q}$ would manifest as the positive gradient of a straight line in a plot of $\ln{F_{q}}$ vs $\ln (1 / \delta \eta)$, which reflects the power law relation between $F_q$ and $\delta \eta$ in equation~\ref{eq:HorizontalMoments}. $\phi_{q}$ been referred to in literature as the ``intermittency exponent'', ``intermittency index'' or ``slope parameter'' and characterises the strength of intermittency~\cite{Lipa:1989yh,Chen:1993fg}.

From its conceptual inception in 1986 up until the late 1990s, intermittency generated great excitement in the high energy physics community. A lot of work was done in its experimental measurements and theoretical studies (see~\cite{DeWolf:1995nyp} for a comprehensive review of the subject), and intermittency seemed to be a universal feature of multiparticle spectra~\cite{Bialas:1998jw}. However, further advances in this field were held up by the lack of data.

With the commissioning of the Large Hadron Collider at CERN in 2009 heralding a new era in TeV collider physics, it is of interest to check for the presence of intermittency at this new energy frontier. In this analysis, we expand on our earlier works~\cite{Ong:2019dlz,Ong:2021hcf} by making an experimental measurement of the intermittency exponents in 1D and 2D for $pp$ collisions at $\sqrt{s}=$ 0.9, 7 and 8 TeV.

\section{About the data}

This analysis is performed on Run 1 data from the CMS collaboration processed from the CMS Open Data Portal, covering centre-of-mass energies $\sqrt{s}=$ 0.9, 7 and 8 TeV. The analysis method follows largely that of CMS~\cite{Khachatryan:2010nk}, which analysed minimum-bias (MinBias), non-single diffractive (NSD) multiplicity distributions.

NSD events were selected by requiring that at least one forward hadron (HF) calorimeter tower on
each side of the detector have at least 3 GeV of energy deposited in the event. The primary vertex was chosen as the vertex with the highest number of associated tracks, which must also be within 15 cm of the reconstructed beamspot in the beam axis and be of good reconstruction quality (ndof $>$ 4).

Good quality tracks were selected by requiring them to carry the \texttt{highPurity} label. Furthermore, we select for tracks with $<$10\% relative error on the transverse momentum ($p_{\text{T}}$) measurement ($\sigma_{p_{\text{T}}} / p_{\text{T}} < 0.1$) to reject low-quality and badly reconstructed tracks. Secondaries were removed by requiring a small impact parameter with respect to the selected primary vertex. Also, tracks were required to have $p_{\text{T}} > 500$ MeV/c, which will be extrapolated to zero via unfolding.

Finally, unfolding was performed using an iterative ``Bayesian unfolding method'', which is more accurately known as ``D’Agostini iteration with early stopping'' and described in~\cite{DAgostini:1994fjx}. This infers the original charged hadron multiplicity distribution (MinBias NSD) from the charged track multiplicity distribution measured.

Tables~\ref{tab:RECOdatasets} and~\ref{tab:MCdatasets} in the Appendix summarise the datasets used.

\section{Results}

In analysing the intermittency data, it was found that excluding $P(0)$ from the analysis results in a clearer intermittency signal (i.e. straight lines with clear positive gradient in the log-log plots). Thus, for each processed multiplicity distribution, $P(0)$ was omitted and the rest of the data points were renormalised to unity. A possible explanation for this effect is the unnaturally high values of $P(0)$ in the unfolded data for $\sqrt{s}=$ 7 and 8 TeV, which would greatly skew the scaling of the horizontal moments (equation~\ref{eq:HorizontalMoments}).

We present our measurement results of the horizontal moments from Run 1 data from the CMS collaboration processed from the CMS Open Data Portal, covering centre-of-mass collision energies $\sqrt{s}=$ 0.9, 7 and 8 TeV. Data treatment has been kept uniform across the energies, so as to observe for trends that appear solely as a function of collision energy. The uncertainties of each data point have been computed and plotted; however, they are small as the partial systematic error analysis has resulted in underestimating the overall error.

\subsection{Intermittency in 1 dimension ($\eta$)}
\label{se:IntermittencyResults1D}

\begin{figure}[h!]
	\centering
	\begin{subfigure}[b]{0.7\textwidth}
		\centering
		\includegraphics[width=\textwidth]{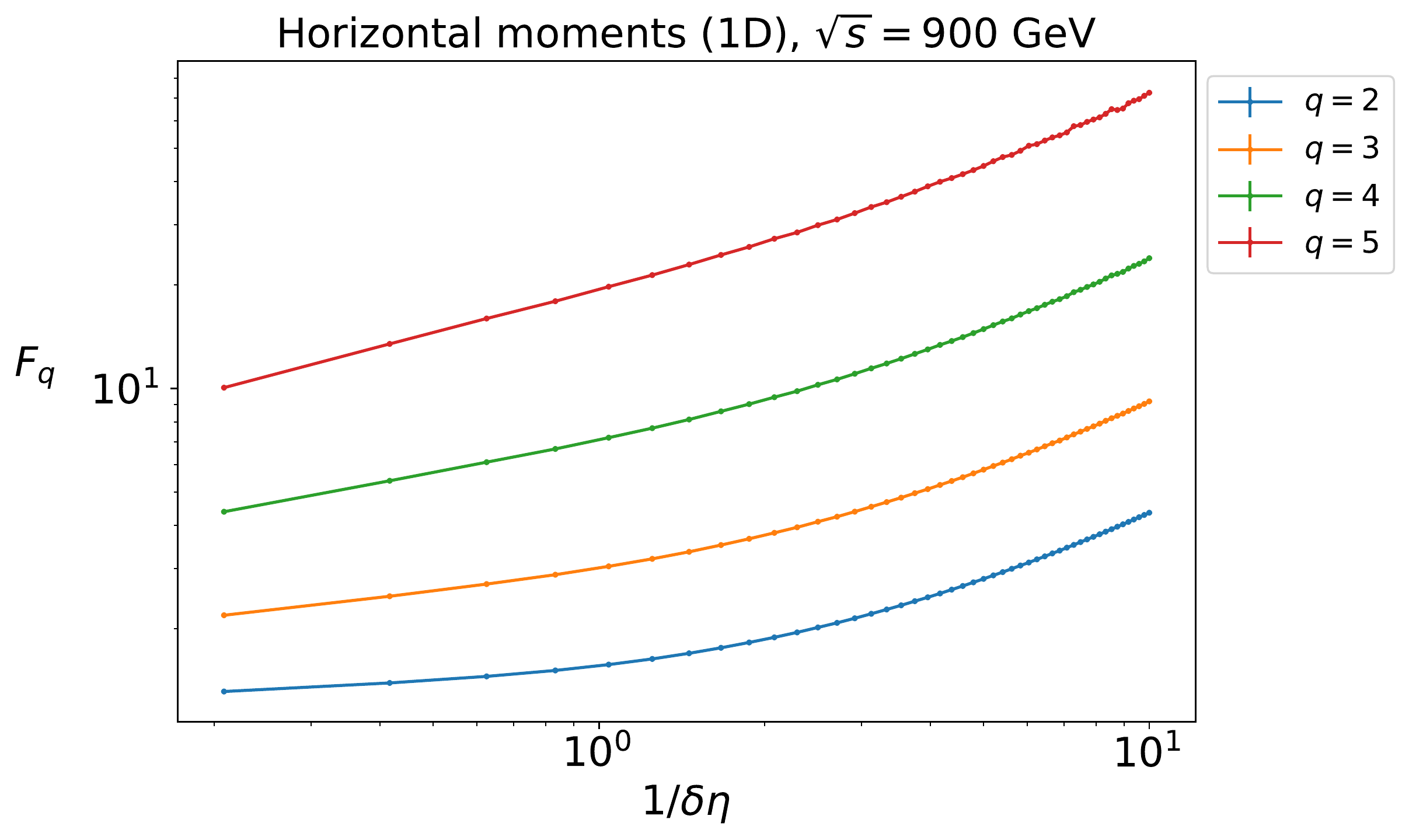}
		\caption{900 GeV}
		\label{fig:Intermittency1D_900GeV}
	\end{subfigure}
	\par\bigskip
	\begin{subfigure}[b]{0.7\textwidth}
		\centering
		\includegraphics[width=\textwidth]{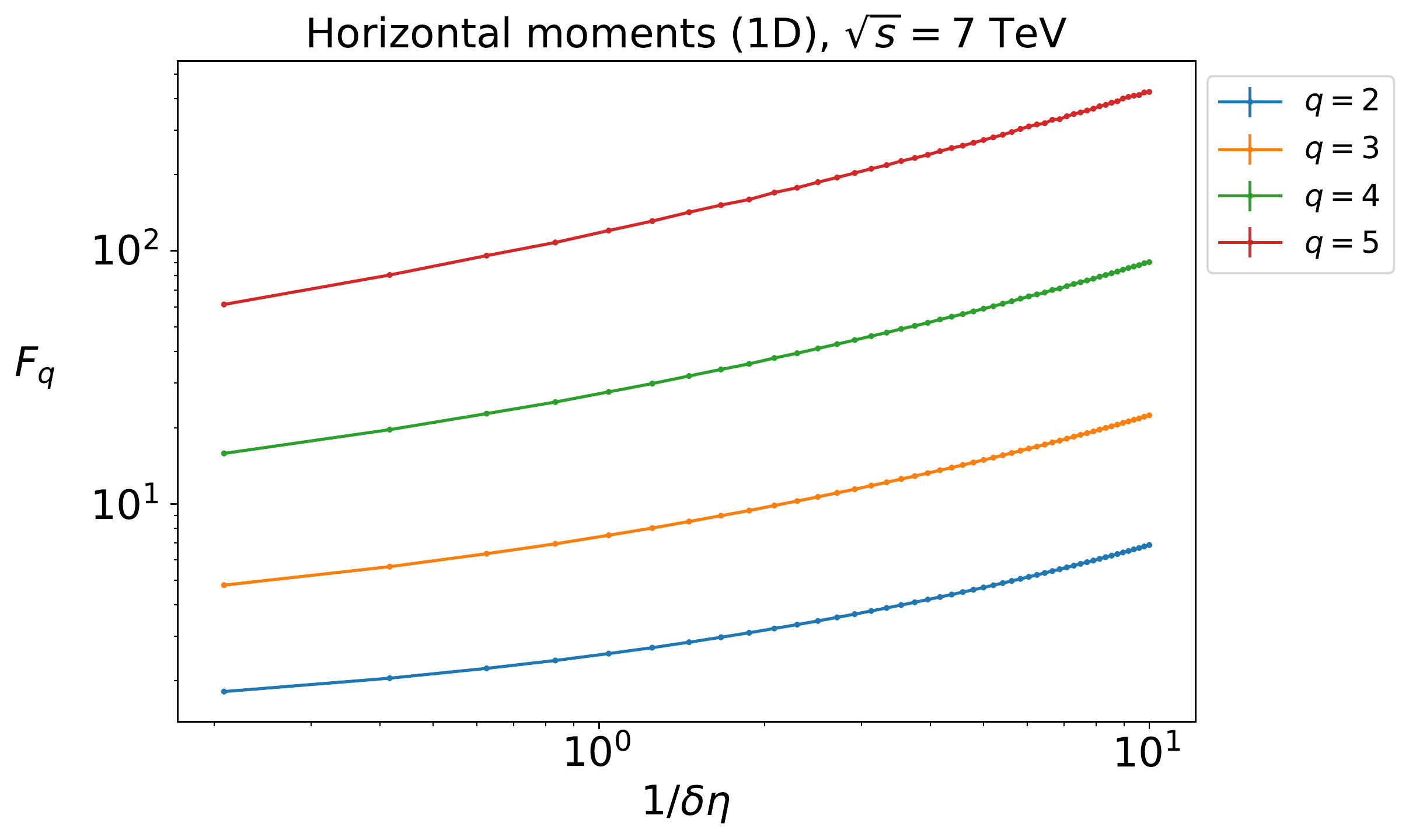}
		\caption{7 TeV}
		\label{fig:Intermittency1D_7eV}
	\end{subfigure}
	\par\bigskip
	\begin{subfigure}[b]{0.7\textwidth}
		\centering
		\includegraphics[width=\textwidth]{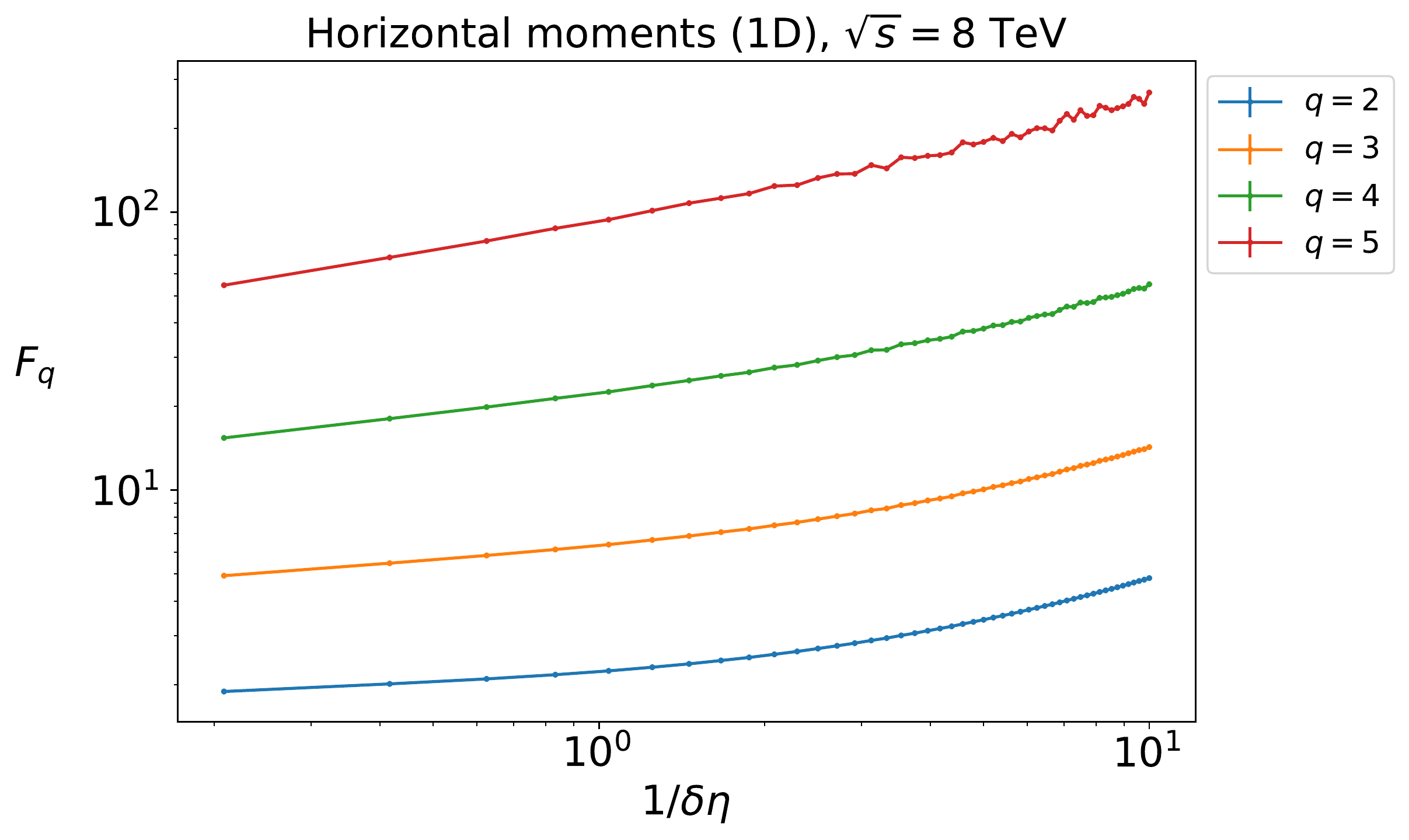}
		\caption{8 TeV}
		\label{fig:Intermittency1D_8GeV}
	\end{subfigure}
	\caption{1-dimensional intermittency -- horizontal moments for $q$ = 2, 3, 4, 5 at (a) $\sqrt{s}=$ 900 GeV, (b) $\sqrt{s}=$ 7 TeV and (c) $\sqrt{s}=$ 8 TeV.}
	\label{fig:Intermittency1D-plots}
\end{figure}

Intermittency has traditionally been studied in the rapidity ($y$) variable. Here, pseudorapidity ($\eta$) will be used instead. In our analysis, we will partition the pseudorapidity space from $M=1$ to $M=48$ bins (and every integer in between), for 48 different partitions with the smallest pseudorapidity resolution at $\delta \eta = 0.1$. Figure~\ref{fig:Intermittency1D-plots} shows the traditional log-log plots, while Table~\ref{tab:Intermittency1D-exponents} presents the intermittency exponents $\phi_q$ experimentally obtained via linear regression on the 10 points corresponding to the smallest $\delta \eta$ sizes.

\begin{table}[h!]
	\centering
	\caption{Summary of 1D intermittency exponents}
	\begin{tabular}{ cccc }
		\hline
		$q$ & 900 GeV & 7 TeV & 8 TeV \\
		\hline
		2	& $ 0.694 \pm 0.002	$ & $ 0.604 \pm 0.003	$ & $ 0.553 \pm 0.004 $ \\
		3	& $ 0.710 \pm 0.005	$ & $ 0.629 \pm 0.003	$ & $ 0.55 \pm 0.01 $ \\ 
		4	& $ 0.73 \pm 0.02	$ & $ 0.649 \pm 0.007	$ & $ 0.54 \pm 0.04 $ \\ 
		5	& $ 0.76 \pm 0.04	$ & $ 0.65 \pm 0.02		$ & $ 0.5 \pm 0.1 $ \\ 
		\hline	
	\end{tabular}	
	\label{tab:Intermittency1D-exponents}
\end{table}

\subsection{Intermittency in 2 dimensions ($\eta$-$\phi$)}
\label{se:IntermittencyResults2D}

\begin{figure}[h!]
	\centering
	\includegraphics[width=.95\linewidth]{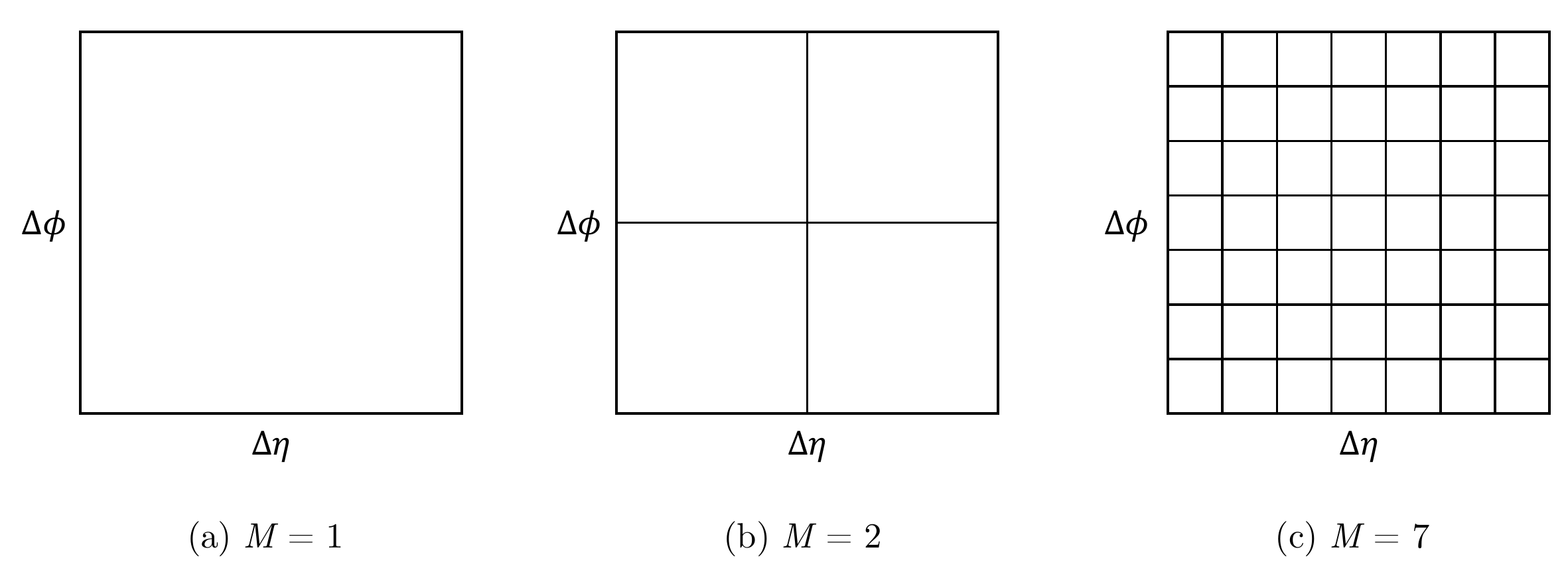}
	\caption{Partition scheme for ($\eta$-$\phi$) phase space in 2D intermittency analysis. The total widths of $\Delta\eta=4.8$ and $\Delta\phi=2\pi$ are each divided into $M$ equal bins, resulting in a total of $M^2$ bins. Shown are (a) $M=1$ for a total of 1 bin, (b) $M=2$ for 4 bins, and (c) maximum value $M=7$ for 49 bins.}
	\label{fig:2DpartitionScheme}
\end{figure}

Intermittency in higher dimensions has also been extensively studied, for example by Ochs~\cite{Ochs:1990wa}. Here, we shall investigate intermittency in ($\eta$-$\phi$), which confers a natural geometric interpretation of the $\alpha$-model particles branching out in 3D-space. We adopt a simple partitioning scheme to carve up the ($\eta$-$\phi$) phase space, with $M=1$ to $M=7$ (and every integer in between): each value of $M$ divides each of the $\eta$ and $\phi$ space into $M$ equal widths, for a total of $M^2$ bins. The scheme is illustrated in Figure~\ref{fig:2DpartitionScheme}.

\begin{figure}[h!]
	\centering
	\begin{subfigure}[b]{0.7\textwidth}
		\centering
		\includegraphics[width=\textwidth]{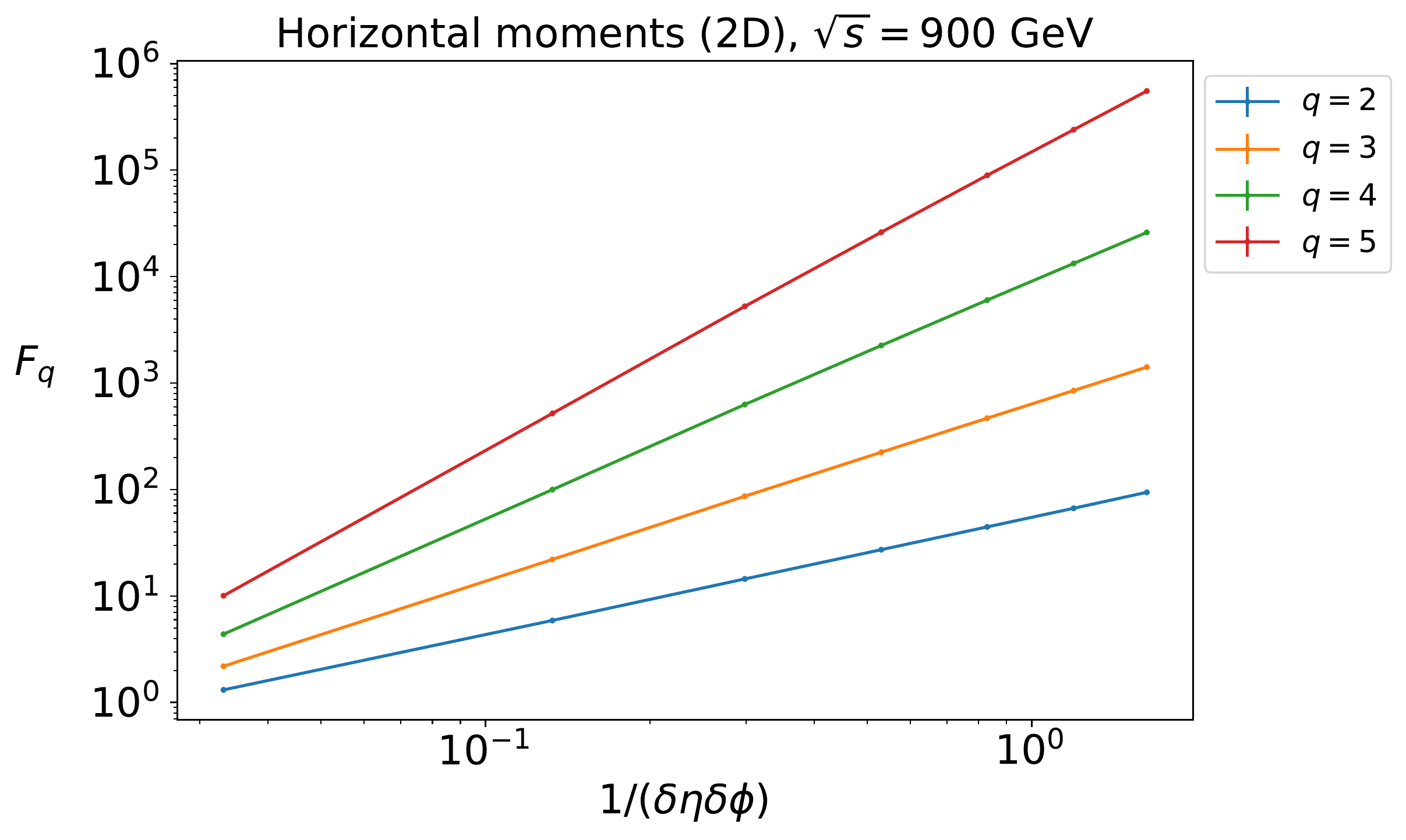}
		\caption{900 GeV}
		\label{fig:Intermittency2D_900GeV}
	\end{subfigure}
	\par\bigskip
	\begin{subfigure}[b]{0.7\textwidth}
		\centering
		\includegraphics[width=\textwidth]{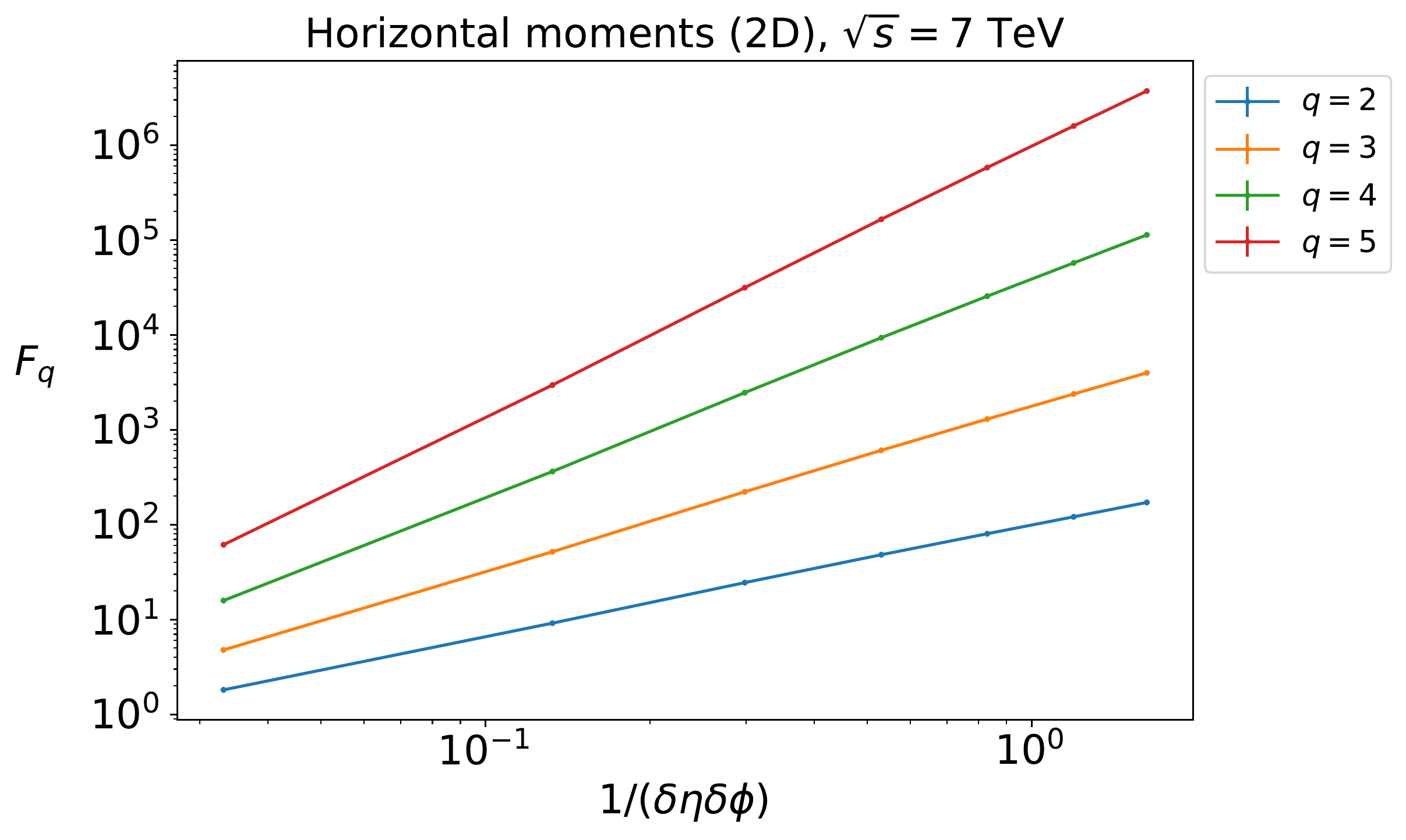}
		\caption{7 TeV}
		\label{fig:Intermittency2D_7eV}
	\end{subfigure}
	\par\bigskip
	\begin{subfigure}[b]{0.7\textwidth}
		\centering
		\includegraphics[width=\textwidth]{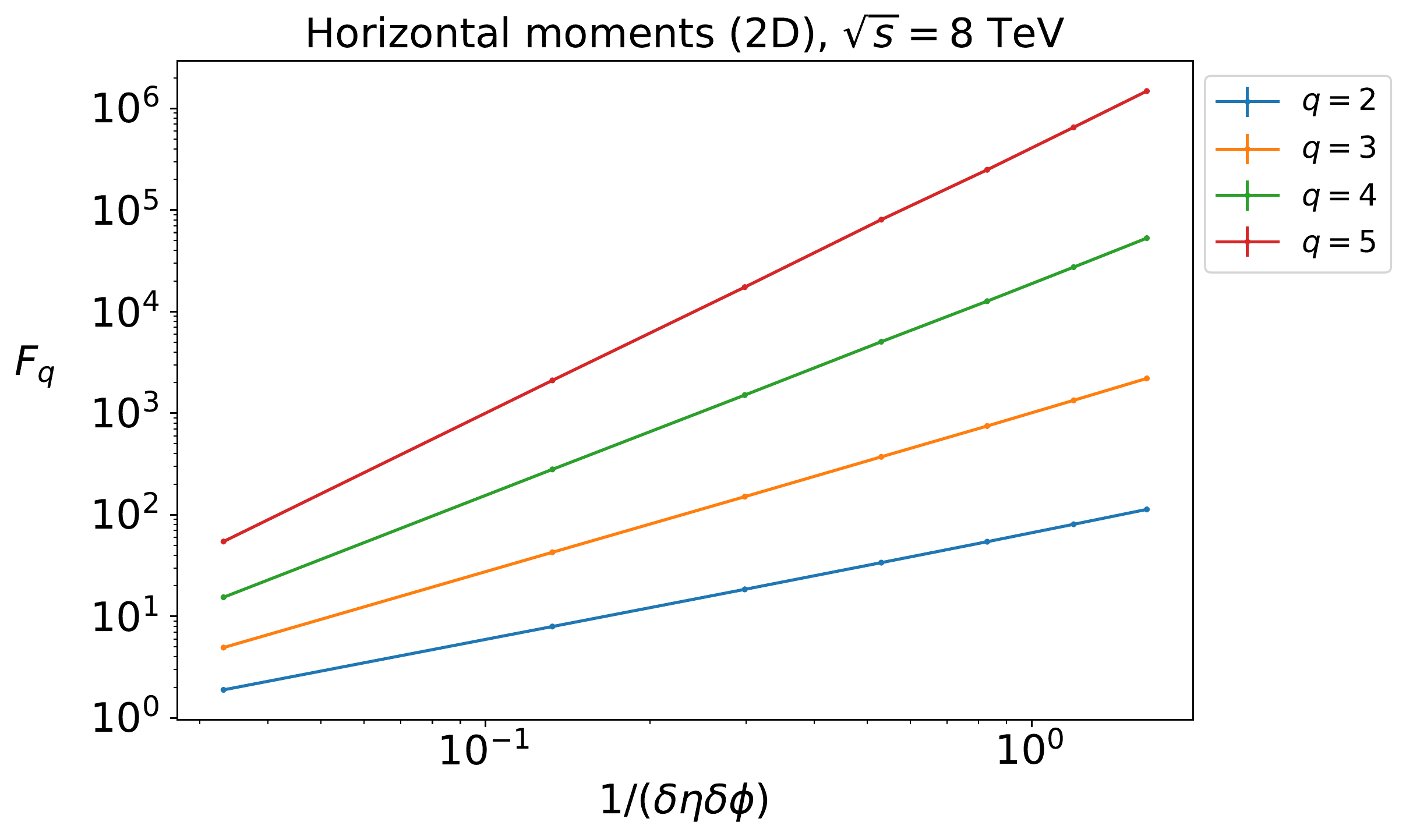}
		\caption{8 TeV}
		\label{fig:Intermittency2D_8GeV}
	\end{subfigure}
	\caption{2-dimensional intermittency -- horizontal moments for $q$ = 2, 3, 4, 5 at (a) $\sqrt{s}=$ 900 GeV, (b) $\sqrt{s}=$ 7 TeV and (c) $\sqrt{s}=$ 8 TeV.}
	\label{fig:Intermittency2D-plots}
\end{figure}

Figure~\ref{fig:Intermittency2D-plots} shows the traditional log-log plots, while Table~\ref{tab:Intermittency2D-exponents} presents the intermittency exponents $\phi_q$ experimentally obtained via linear regression on the 5 points corresponding to the smallest $\delta \eta \times \delta \phi$ sizes.

\begin{table}[h!]
	\centering
	\caption{Summary of 2D intermittency exponents}
	\begin{tabular}{ cccc }
		\hline
		$q$ & 900 GeV & 7 TeV & 8 TeV \\
		\hline
		2	& $ 1.105 \pm 0.002	$ & $ 1.147 \pm 0.006 $ & $ 1.069 \pm 0.005 $  \\
		3	& $ 1.648 \pm 0.002 $ & $ 1.70 \pm 0.01	  $ & $ 1.580 \pm 0.005 $  \\ 
		4	& $ 2.195 \pm 0.006 $ & $ 2.26 \pm 0.01   $ & $ 2.095 \pm 0.007 $  \\ 
		5	& $ 2.75 \pm 0.01	$ & $ 2.81 \pm 0.02	  $ & $ 2.61 \pm 0.01 $  \\ 
		\hline	
	\end{tabular}	
	\label{tab:Intermittency2D-exponents}
\end{table}

\section{Discussion and conclusion}

Table~\ref{tab:Intermittency1D-exponents} shows that the intermittency strength in 1D decreases with increasing energy. This is a possible indication that the $\alpha$-model of multiparticle production is getting suppressed.

For the 2D case, Table~\ref{tab:Intermittency2D-exponents} shows that the intermittency exponents for 2D have a greater magnitude than the 1D case, indicating a stronger intermittency signal. However, the trends with increasing energy is not observed, as the exponents rise slightly from 900 GeV to 7 TeV before a dropping quite significantly at 8 TeV. When compared to the results from 1D analysis, the 2D exponents for 7 TeV are most likely to be anomalous, and would require further investigation.

The results of this chapter strongly indicate that intermittency is still present in multiparticle production in $pp$ collisions up to $\sqrt{s}=$ 8 TeV. This suggests that multiparticle production mechanisms, at energies up to the TeV scale, ought to have features such as self-similarity, scaling, correlations and branching.

\clearpage
\appendix

\section{Datasets used}

\begin{table}[h!]
	\centering
	\caption{Summary of CMS collider datasets used from CERN Open Data Portal}
	\begin{tabular}{ |c|l|c| } 
		\hline
		$\sqrt{s}$  & \multirow{2}{*}{Dataset} & \multirow{2}{*}{Ref.} \\
		(TeV) & & \\
		\hline \hline
		0.9	& /MinimumBias/Commissioning10-07JunReReco\_900GeV/RECO & \cite{data900GeV} \\
		\hline
		7	& /MinimumBias/Run2010A-Apr21ReReco-v1/AOD & \cite{data7TeV} \\ 
		\hline
		8	& /MinimumBias/Run2012B-22Jan2013-v1/AOD & \cite{data8TeV} \\ 
		\hline	
	\end{tabular}	
	\label{tab:RECOdatasets}
\end{table}

\begin{table}[h!]
	\centering
	\caption{Summary of Monte Carlo datasets used from CERN Open Data Portal}
	\begin{tabular}{ |c|l|c| } 
		\hline
		$\sqrt{s}$  & \multirow{2}{*}{Dataset} & \multirow{2}{*}{Ref.} \\
		(TeV) & & \\
		\hline \hline
		
		\multirow{2}{*}{0.9}	& /MinBias\_TuneZ2\_900GeV\_pythia6\_cff\_py & \multirow{2}{*}{\cite{mc900GeV}}\\
		& \_GEN\_SIM\_START311\_V2\_Dec11\_v2 &  \\
		\hline
		
		\multirow{2}{*}{7}	& /MinBias\_TuneZ2star\_7TeV\_pythia6/Summer12-LowPU2010 & \multirow{2}{*}{\cite{mc7TeV}} \\ 
		& \_DR42-PU\_S0\_START42\_V17B-v1/AODSIM &  \\ 
		\hline
		
		\multirow{2}{*}{8}	& /MinBias\_TuneZ2star\_8TeV-pythia6/Summer12\_DR53X-PU & \multirow{2}{*}{\cite{mc8TeV}} \\
		& \_S10\_START53\_V7A-v1/AODSIM & \\
		\hline	
	\end{tabular}	
	\label{tab:MCdatasets}
\end{table}

\printbibliography

\end{document}